\documentclass[aps,prd,reprint,superscriptaddress,nofootinbib]{revtex4-1}
\usepackage{cuted} 
\usepackage[normalem]{ulem}
\usepackage[normalem]{ulem}
\usepackage{xcolor}
\usepackage{mdframed}
\usepackage[T1]{fontenc} 

\usepackage{graphicx,epsfig}
\usepackage{bbm}
\usepackage{graphics,dcolumn,bm, float}
\usepackage{amssymb,amsmath,rotate,color,amsfonts}
\usepackage[title,titletoc,toc]{appendix}
\usepackage{mathtools}
\usepackage{booktabs}
\usepackage{tcolorbox}
\usepackage{tikz-cd}
\usetikzlibrary{shadings}
\usepackage{xcolor}
\usepackage{tikz-3dplot}
\usepackage{setspace}
\usepackage{wrapfig}
\usepackage{lipsum}
\usepackage{mwe}
\usepackage[mathlines]{lineno}
\usepackage[mathscr]{euscript}
\usepackage{bbold}
\usepackage{pgfplots}
\usepackage{float}
\usepackage{tkz-euclide}
\usepackage{braket}
\usepackage{physics}
\usepackage[caption=false]{subfig}
\usepackage[export]{adjustbox}
\usepackage{tikz}
\usepackage{slashed}
\usepackage{bm}
\usepackage{multirow}
\usetikzlibrary{through,calc}
\usetikzlibrary{positioning}

\usepackage{hyperref}

\newcommand{\be}{\begin{equation}}
	\newcommand{\ee}{\end{equation}}
\newcommand{\bea}{\begin{eqnarray}}
	\newcommand{\eea}{\end{eqnarray}}

\newcolumntype{P}[1]{>{\centering\arraybackslash}p{#1}}
\tdplotsetmaincoords{70}{120} 

\newtheorem{conjecture}{Conjecture}

\begin{document}

\title{The species scale and the refined TCC bounds}

\preprint{}

\author{Ahmad Moradpouri}
\affiliation{Research Center for High Energy Physics, Department of Physics, Sharif University of Technology, P.O.Box 11155-9161, Tehran, Iran.}
\thanks{Emails:Ahmadreza.Moradpour@gmail.com}
\begin{abstract}
The species scale $\Lambda_s$, is the energy scale at which quantum corrections to Einstein’s theory of gravity become significant. In many cases, this scale corresponds to the string mass scale, or the higher-dimensional Planck scale, which can be much lower than the Planck scale in the weak coupling limit. In this note, we explore two variants of the trans-Planckian censorship conjecture (TCC), which we refer to as the refined TCC conjectures. These versions are related to the species scale as a new cutoff in quantum gravity, and we study whether stringy or quantum modes remain stringy or quantum and never exceed the Hubble horizon. The refined TCCs predict a shorter lifetime for de-Sitter spacetime and impose a bound on the cosmological constant. We also show that one of these conjectures violates some decelerating expansions as opposed to the standard TCC bound. We present supporting evidence for the refined TCCs where its consequences are compatible with other swampland results. The most clear consequence of it is the lower bound for $|\frac{\Lambda'_s}{\Lambda_s}|>\frac{1}{\sqrt{(d-1)(d-2)}}$ which is expected from other swampland criteria. Furthermore, it is argued that both the TCC and the refined TCC are incompatible with the big-bang singularity. Therefore, if the TCC and the refined TCC hold, the big-bang singularity must be ruled out. In scenarios such as string gas cosmology, where the TCC is realized, the spacetime background remains completely regular. In addition to this line of reasoning, we argue for a lower bound on the $\frac{d(1/H)}{dt}$ which excludes the de-Sitter spacetime, which is a regular spacetime. 
\end{abstract}

	\maketitle
	\flushbottom
	
	
	
	
	\section{Introduction}
String theory is currently the leading candidate for explaining the principles of quantum gravity. However, from a phenomenological standpoint, its biggest challenge lies in providing concrete, testable predictions. The Swampland program aims to address this by proposing conjectures that are believed to be satisfied by all consistent effective field theories when coupled with gravity~\cite{palti2019swampland,VANBEEST20221,agmon2022lectures}. These conjectures impose nontrivial constraints on cosmological and particle physics models which have many consequences from early universe, dark energy and dark matter to phenomenological particle physics~\cite{Ooguri_2019,bedroya2020trans,Montero_2023,Gendler:2024gdo,Obied:2023clp,Law-Smith:2023czn,Gonzalo:2022jac,Bedroya:2019tba,Gonzalo:2021zsp,Vafa:2024fpx}.

The conventional energy scale at which quantum corrections to Einstein's theory of gravity are expected to become significant is the Planck mass scale. However, as is argued in~\cite{Dvali_2010,dvalispecies,Dvaliblackholes,Dvali_2013}, the species scale, $\Lambda_s$, introduces a lower cutoff in presence of large number of light particles. Based on the recent investigations~\cite{van_de_Heisteeg_2024,vandeheisteeg2023boundsspeciesscaledistance,vande}, it is suggested that the species scales must be identified by a scale where the higher derivative corrections to the Einstein gravity should be accounted. 

On the other hand, the emergent string conjecture~\cite{lee2022}, characterizes the nature of the tower of light particles which are connected to the species scale. Therefore, in the infinite distance of the moduli space of the theory, either we have a decompactified phase or a weakly coupled string theory. Then the species scale corresponds to the higher-dimensional Planck mass or the string mass scale. It is interesting to explore how the species scale affects other conjectures in the swampland program.

The trans-Planckian conjecture~\cite{bedroya2020trans} is one of the key cosmological conjectures, placing bounds on the rate of the universe's expansion scale factor. More precisely, it states the following:

\bea
\label{TCC}
\frac{a_f}{a_i}l_P<\frac{1}{H_f},
\eea
It states that sub-Planckian modes should remain quantum and never exceed the size of the Hubble horizon. Under time reversal, the Trans-Planckian Censorship Conjecture (TCC) applies to contracting universes, taking the following form:
\bea
\label{timeinversionTCC}
\frac{a_f}{a_i}(-\frac{1}{H_i})>l_P
\eea
The contracting case can be interpreted through the lens of the "no big bang singularity" conjecture. For a given initial Hubble horizon $-\frac{1}{H_i}$, if the theory predicts a singularity, the final scale factor approaches zero, $a_f\to 0$, violating the TCC inequality. Thus, the TCC suggests that the contracting phase of the universe cannot lead to a singularity. Even at the classical level, the singularity theorems (such as~\cite{penrose-singularity-theorem,Hawking:1967ju,Hawking:1970zqf,PhysRevLett.90.151301} and for more examples see~\cite{Senovilla_2015,Steinbauer_2022} and~\cite{Senovilla_1998} for a review) indicate that general relativity is incomplete, necessitating a more fundamental theory to resolve this issue.

Since the species scale is the true scale at which quantum corrections to the classical theory must be considered, and can in principle be much smaller than the Planck scale, it is interesting to examine how the TCC behaves under this new cutoff. The relation between the species scale and the TCC conjecture is explored in~\cite{bedroya2025speciesscaledrivenbreakdowneffective}. Some versions of the refined TCC have also been studied; for example, \cite{Cai_2020} employs the scalar weak gravity conjecture. In these refined TCC conjectures, the naive Planck length is replaced by the species scale $\Lambda_s=\frac{1}{l_{\Lambda_s}}$, where $l_{\Lambda_s}$  is the characteristic length associated with it. Consequently, the first proposed refined TCC states that modes with wavelength $\lambda\sim l_{\Lambda_s}$  never exceed the Hubble horizon, as follows:

\begin{conjecture}
\bea
\label{conjecture1}
\frac{a_f}{a_i}l_{\Lambda_s(t_i)}<\frac{1}{H_f}
\eea
\end{conjecture}

Nevertheless, the refined conjecture differs from the standard TCC in a number of important respects. The first is that the species length scale $l_{\Lambda_s}$ is no longer a fixed parameter but is allowed to evolve in time. Such temporal variation can modify the lifetime of de Sitter spacetime, yielding a value that may be parametrically smaller than the standard TCC bound of $\frac{1}{H}Log(\frac{1}{H})$.

The second major difference is that, according to the string emergent conjecture, the species scale can coincide with the string mass scale. Consequently, in certain scenarios, stringy modes must also be prohibited from exceeding the Hubble horizon. The string length is determined by the string coupling constant, and when the theory is weakly coupled, this length can become large. In such situations, we may impose a stricter bound not only on the lifetime of de Sitter spacetime but also on the value of the cosmological constant itself. Nevertheless, it is possible to put forward a stronger version of the refined TCC conjecture, which can be stated as follows:
\begin{conjecture}
	\label{conjecture2}
	\bea
	\frac{a_f}{a_i}l_{\Lambda_s(t_i)}\leq l_{\Lambda_s(t_f)}
	\eea
\end{conjecture}

Given that the species scale varies across the moduli space, the refined conjecture effectively compares the cutoff scales at different moduli-space points in a cosmological evolution. Specifically, it requires that the cutoff at $t=t_f$ exceeds that at $t=t_i$, despite the expansion that takes place in between. To see the reasoning, consider a quantum mode with initial wavelength $\lambda_i\sim\Lambda_s(t_i)$ at time $t_i$. Under cosmological expansion, this wavelength grows and will ultimately surpass the species scale at the later time $t=t_f$
\bea
\lambda_f=\frac{a_f}{a_i}\lambda_i>l_{\Lambda_s(t_f)},
\eea 

Since this mode is larger than the species scale at $t=t_f$, there is nothing inconsistent with allowing it to become classical at later times $t>t_f$. However, when compared with the initial time $t=t_i$, we conclude that the quantum mode becomes classical only at $t>t_f$, thereby violating the essence of the standard TCC conjecture. We therefore assume the above refined conjecture to hold.

The organization of the paper is as follows: After introducing the refined TCC bounds in the introduction, in Section~\ref{ref1} we study the consequences of these conjectures, including bounds on the cosmological constant and the de Sitter lifetime. In Section~\ref{ref2}, we investigate the relation between the standard and the refined TCC bounds and the no singularity conjecture and argue that both of them rule out the singularity of the Big Bang at the beginning of time. Furthermore, we rationalize a new lower bound on $\frac{d(1/H)}{dt}$ in the view of the no singularity conjecture, which excludes de Sitter spacetime. In Section~\ref{species scale slope}, we also show that the second refined TCC bound~\ref{conjecture2} has a clear consequence on the slope of the species scale $|\frac{\Lambda_s'}{\Lambda_s}|$ which is compatible with other swampland conjectures.

\subsection{The dark energy:  the TCC vs the refined TCCs}
\label{ref1}
As mentioned in the introduction, the TCC requires that for any expanding universe, the cosmological scale factor and the Hubble parameter over the time interval $[t_i,t_f]$ must satisfy Eq.~\eqref{TCC}. While most discussions of the conjecture assume that the initial and final times, $t_i$ and $t_f$, are distinct, here we explore the special case where the initial and final times coincide, $t_i = t_f = t$,
\bea
\label{equal-TCC}
l_P<\frac{1}{H(t)},
\eea
which asserts that the Hubble horizon must always be greater than the Planck length at all times.

The Hubble parameter in the de-Sitter spacetime is constant, imposing specific constraints on its value,
\bea
\label{H-bound}
H<\frac{1}{l_P},
\eea
It is not surprising that the cosmological constant should not be too large, as the structure formation and the anthropic principle impose constraints on its value~\cite{PhysRevLett.59.2607}. Although Eq.~\eqref{H-bound} is, in some sense, trivial, reflecting the requirement that the energy scales remain below the Planck scale, it offers a more attractive and fundamental perspective. It also opens up scenarios in which dark energy can be analyzed. In particular, we will test it in the context of the refined TCC, which provides more nontrivial constraints on the dark energy.

Let us now examine the refined TCC and explore how it constrains the Hubble constant. Since the emergent string conjecture claims that the light towers at infinite distance correspond to either a decompactification phase or a weakly coupled string theory, let us for a moment assume that the string mass scale coincides with the species scale. Since all swampland conjectures are expressed in the Einstein frame, in order to state the refined TCC, we need to measure the string length in the Einstein frame as well. Using the appropriate Weyl transformation in this way,
\bea
g_{\mu\nu}^{E}\equiv e^{-\frac{4\phi}{d-2}}g_{\mu\nu}^S
\eea
where $\phi$ is the dilaton field, we can relate the lengths in two frames. The Planck length $l_p$ is related to the string length as follows
\bea
l_p=e^{\frac{2\phi}{d-2}}l_s,
\eea
where in the unit of the Planck length the first conjecture~\ref{conjecture1} takes the following form

\bea
\label{refinedtcc2}
\frac{a_f}{a_i}g_s(t_i)^{-\frac{2}{d-2}}<\frac{1}{H_f}.
\eea
where $g_s=e^{\phi}$ is the string coupling. The TCC places an upper bound on the lifetime of de-Sitter spacetime as $T\leq\frac{1}{H}log\frac{M_p}{H}$. Needless to say, as the species scale can be smaller than the naive Planck length, the refined TCC predicts a shorter lifetime for de-Sitter spacetime, which may align more coherently with other swampland conjectures, such as the de-Sitter conjecture, which rules out de-Sitter spacetime.  Another interesting point of the refined TCC~\eqref{conjecture1} is that it imposes a more nontrivial constraint on the cosmological constant itself.  For $t_i=t_f$, we conclude that : 

\bea
l_P g_s(t)^{-\frac{2}{d-2}}<\frac{1}{H(t)}.
\eea
For the de-Sitter spacetime with a constant Hubble parameter $H$, the following condition holds for all times:

\bea
\label{darkenergybound}
H<\frac{g_s(t)^{\frac{2}{d-2}}}{l_P}.
\eea 
The cosmological constant $\Lambda$, and the Hubble parameter $H$, are related by $H^2\sim\Lambda$. Therefore, the refined $TCC$ put an upper bound on the cosmological constant.

Equation~\eqref{darkenergybound} may explain "why the cosmological constant is so small compared to the Planck energy scale, $M_P^2$." The answer may lie in the fact that we live in an approximately free limit of string theory, where string interactions are not too large. However, it cannot solve the hierarchy of scales, as the species scales cannot be smaller than the energy scales, smaller than the TeV scale, where the standard model of particle physics, as an effective field theory give an accurate description of nature.

There are some straightforward consequences of conjectures~\ref{conjecture1} and~\ref{conjecture2} that are distinct from the standard TCC bound.

\begin{itemize}
	\item 
	Since $\Lambda_s < 1$ in Planck units, the first conjecture~\ref{conjecture1} implies a shorter lifetime for de Sitter spacetime than the standard TCC. However, a more interesting prediction comes from the second conjecture, which may further shorten the de Sitter lifetime. Given the natural assumption that the UV cutoff is greater than the IR cutoff, i.e., $\Lambda_s > H$, conjecture~\ref{conjecture2} directly implies a shorter lifetime for de Sitter spacetime:
	\begin{equation}
		\label{shorter_lifetime}
		\Delta T \leq \frac{1}{H_f} \log \frac{\Lambda_s(t_i)}{\Lambda_s(t_f)} < \frac{1}{H_f} \log\left(\frac{\Lambda_s(t_i)}{H_f}\right).
	\end{equation}
	
	\item 
	Similar to the standard TCC, conjecture~\ref{conjecture1} requires accelerating expansion in order to be violated, provided that the Hubble parameter is always smaller than the species scale, $H < \Lambda_s$. However, a striking feature of conjecture~\ref{conjecture2} is that it can also exclude certain decelerating expansions. Suppose conjecture~\ref{conjecture2} is violated; then we have
	\begin{equation}
		\label{violation_condition}
		a_f \Lambda_s(t_f) > a_i \Lambda_s(t_i).
	\end{equation}
	Since $H < \Lambda_s$, we have $\dot a < a \Lambda_s$. However, one cannot always deduce that
	\begin{equation}
		\label{velocity_inequality}
		\dot a_f > \dot a_i,
	\end{equation}
	and it is therefore possible for some decelerating expansions to also violate conjecture~\ref{conjecture2} which are generally consistent with the standard TCC bound.
\end{itemize}

\subsection{How to see resolving the big bang singularity in quantum gravity?}
\label{ref2}
As mentioned in the introduction, the singularity theorems imply that general relativity, even at the classical level, is incomplete. Addressing this issue requires a deeper theory. A true quantum theory of gravity should provide a resolution to this problem. Two possible approaches to reformulating the answer are as follows: 

\begin{itemize}
	\item Negative perspective: Resolving the issue by dissolving the problem, i.e., reinterpreting or reframing the singularities as artifacts of an incomplete framework rather than fundamental physical realities. 
\end{itemize} 
In such scenarios, which are the most plausible within a final quantum theory of gravity, the true degrees of freedom are not geometric, as in Riemannian geometry. Instead, singularities emerge as large-scale features of an emergent metric. In this framework, discussing lengths around the Planck scale becomes meaningless, and the problem of singularities naturally dissolves within this picture.

\begin{itemize}
	\item Positive perspective: Addressing the problem directly by finding solutions within the framework of an effective field theory for gravity, where singularities are resolved by some higher derivative or matter terms. 
\end{itemize}
From this perspective, although the theory of quantum gravity might not be formulated using geometric ingredients, any UV completion of general relativity should admit an effective action expressed in geometric terms. We require that this action prohibit any singular metric as a solution to its equations of motion.

String theory's dualities, such as $AdS/CFT$, $T$-duality, and mirror symmetry, suggest that duality-invariant quantities are not inherently geometric. This points to a deeper, yet-to-be-discovered mathematical structure underlying all currently known concepts, which represents the true mathematical foundation of string theory. In certain extreme limits of the moduli space parameters, this structure reduces to the familiar string theories. From this perspective, string theory may align with the negative approach, dissolving the problem of singularities by rendering them artifacts of an emergent description.

Nevertheless, in certain extreme limits of the parameter space, examples of string theory can still be formulated in terms of geometric concepts and effective actions. This holds for all known examples of string theory, at least to some order in the string length $l_s$ and the string coupling constant $g_s$. By turning off all non-geometric fields, such as the Kalb-Ramond field and others, and assuming a constant dilaton field $\phi$, the tree-level effective action takes a higher derivative expansion in terms of $l_s$ as follows 

\bea
\label{effectiveaction}
S\sim\int d^{d+1}x\sqrt{-g}e^{-2\phi}\Big(\mathcal{R}+\lambda l_s^2\mathcal{R}_{\mu\nu\rho\sigma}\mathcal{R}^{\mu\nu\rho\sigma}+...\Big),
\eea
where $\lambda$ is a different real number for different string theories. From this perspective, we may expect that string theory resolves at least some of the most challenging singularities, such as the big-bang or those arising in black holes.\footnote{String theory may permit certain types of singularities, such as orbifold singularities, which are well understood and controllable. However, in this paper, we focus exclusively on the big bang singularity and assume that string theory resolves it.} This view aligns with the positive perspective in the perspectives. For the remainder of this paper, we adopt this view, and whenever we refer to string theory resolving a singularity, it should be understood in light of this framework.
\subsubsection{Big bang singularity: the TCC vs the refined TCC}
\label{bigbangsingularity}
Adopting this interpretation of resolving the big bang singularity, the effective actions of string theories, to all orders in $\alpha'$ and $g_s$, share a similar nature as~\eqref{effectiveaction}, with the inclusion of additional matter fields. Based on the TCC, all expanding universes derived from this action must satisfy the TCC bound.  Let us examine whether power-law expansions, $a(t)\sim t^{\alpha}$, are compatible with the TCC and the refined TCC or not. The TCC-bound Eq.~\eqref{equal-TCC} becomes particularly interesting, as the Hubble horizon grows linearly with time  
\bea
\frac{1}{H}\sim t,
\eea
and Eq.~\eqref{equal-TCC} then leads to the following inequality 

\bea
l_P<t,
\eea
near the big bang singularity at $t\sim 0$, the length of the Hubble horizon can become smaller than the Planck length, violating the bound Eq.~\eqref{equal-TCC}. Indeed, the relation Eq.~\eqref{equal-TCC} supports the no-singularity conjecture, suggesting that the big bang singularity should be resolved by some mechanism, possibly through the inclusion of higher-curvature and loop corrections in string theory to satisfy the bound Eq.~\eqref{equal-TCC}.

Let us focus on the curvature invariants. One can give a simple argument for the no-singularity conjecture.  For example, the expansion scalar $\theta$, has a clean geometrical meaning which quantifies the fractional rate of change of volume~(area) $\frac{1}{\delta V}\frac{d\delta V}{d\tau}$($\frac{1}{\delta A}\frac{d\delta A}{d\tau}$) for time-like~(null-like) geodesics. The Raychaudhuri's equation for a congruence of time-like geodesics is given by~\cite{Poisson:2009pwt}, 
\bea
\frac{d\theta}{d\tau} = -\frac{1}{3} \theta^2 - \sigma^{\alpha\beta} \sigma_{\alpha\beta} + \omega^{\alpha\beta} \omega_{\alpha\beta} - R^{\alpha\beta} u_\alpha u_\beta,
\eea
is the fundamental equation in exploring the singularity theorems and shows the importance of $\theta$ in analyzing singularities. Let us consider a congruence of time-like comoving world lines in an expanding universe with the following metric in $d=3+1$ spacetime
\bea
ds^2=-dt^2+a^2(t)d\vec x^2,
\eea
and the expansion scalar for the tangent vectors $u_{\alpha}=-\partial_{\alpha}t$, is given by
\bea
\theta=\frac{1}{a^3}\frac{da^3}{dt}=3H,
\eea
and when the TCC satisfies, Eq.\eqref{H-bound} implies that $\frac{1}{3}|\theta|<\frac{1}{l_P}$. Therefore, for such spacetimes, a caustic point ($|\theta|=\infty$) cannot form. The finiteness of curvature invariants also demands that $\dot H$ should be bounded. For example, the Kretschmann invariant for the FRW metric is given by
\begin{equation}
	K = R_{\mu\nu\rho\sigma}R^{\mu\nu\rho\sigma} 
	= 12 \left( (\dot H + H^2)^2 + \left( \frac{k}{a^2} + H^2 \right)^2 \right),
	\label{kretschmann}
\end{equation}
and the boundedness of the Kretschmann invariant implies that $\dot H$ is also bounded. The no-singularity conjecture is not strong enough to exclude de Sitter spacetime, which is perfectly regular, and one cannot put a lower bound on $\dot H$(or equivalently on $\frac{d(\frac{1}{H})}{dt}$) as $\dot H>\alpha$ for a positive number $\alpha$. However, following conjecture~\ref{conjecture2}, we may suggest a bound that could potentially exclude de Sitter spacetime. 

\begin{itemize}
	\item \textbf{First argument}: 
 \end{itemize}    
    
Taking the infinitesimal limit of conjecture~\ref{conjecture2}, with $t_f = t_i + \delta t$, and working in Planck units, we obtain
\begin{equation}
	\label{infinitesimal1}
	\frac{d l_{\Lambda_s}}{dt} \, \frac{\Lambda_s}{H} > 1.
\end{equation}
Since in Planck units $\Lambda_s < 1$ always holds, this simplifies to
\begin{equation}
	\label{infinitesimal2}
	\frac{d l_{\Lambda_s}}{dt} \, \frac{1}{H} > 1.
\end{equation}
Assuming the effective field theory description remains valid, the inequality
\begin{equation}
	\label{inequality}
	\frac{d l_{\Lambda_s}}{dt} > \frac{d}{dt}\left(\frac{1}{H}\right)
\end{equation}
cannot hold for all times; otherwise, the species length scale $l_{\Lambda_s} = \Lambda_s^{-1}$ would grow to exceed the Hubble horizon at sufficiently late times. Consequently, we arrive at the following bound, which must hold for nearly all times:
\begin{equation}
	\label{final_bound}
	\frac{d}{dt}\left(\frac{1}{H}\right) \frac{1}{H} > 1.
\end{equation}
The time scale over which inequality~\eqref{inequality} is violated is related to the lifetime of de Sitter spacetime. However, since some swampland conjectures, such as the de Sitter conjecture, exclude de Sitter spacetime entirely, even a metastable one, it is instructive to examine the consequences of bound~\eqref{final_bound} under the assumption that it holds at all times. The bound is trivially satisfied during radiation- and matter-dominated eras, while it excludes de Sitter spacetime, since $\frac{1}{H}$ is constant and does not vary with time. The above bound places a lower bound on the rate of change of the Hubble horizon for expanding universes:
\begin{equation}
	\label{lower_bound}
	\frac{d}{dt}\left(\frac{1}{H}\right) > H.
\end{equation}
However, the above bound does not exclude accelerating expansion, as we may have $\ddot a>0$.

 It is also straightforward to derive an upper bound on $\frac{d}{dt}(1/H)$ directly from the Einstein equations. Assuming that the Einstein equations in a cosmological setting are reliable for a single scalar field where we assume to be positive, in reduced Planck mass units we have
\begin{align}
	\frac{(d-1)(d-2)H^2}{2} &= \frac{1}{2}\dot\phi^2 + V, \label{einstein1} \\
	(d-2)\dot H &= -\dot\phi^2, \label{einstein2}
\end{align}
which leads to the following inequality:
\begin{equation}
	\label{upper_bound}
	\frac{d}{dt}\left(\frac{1}{H}\right) \leq (d-1).
\end{equation}

Nevertheless, relation~\eqref{final_bound} does not yield a well-motivated constraint on the potential slope $\left|V'/V\right|$. By contrast, if we adopt a slightly strengthened version of~\eqref{lower_bound}, the resulting bound on the potential slope assumes a familiar form and aligns more naturally with other swampland conjectures. The bound is just that there is no accelerating expansion:
\begin{equation}
	\label{stronger_bound}
	\frac{d}{dt}\left(\frac{1}{H}\right) \geq 1.
\end{equation}
Since~\eqref{lower_bound} follows directly from this inequality(since we always have $H<1$), the proposed bound is manifestly stronger than~\eqref{lower_bound}. The Einstein equations take the following form:
\bea
&&V(\phi)=\Big(\frac{(d-1)(d-2)}{2}H^2-\frac{(d-2)^2}{2}H^{'2}\Big),\\\label{fried2}
&&(d-2)H'=-\dot\phi,
\eea
where $'$ denotes derivative with respect to $\phi$. From bound~\eqref{stronger_bound}, we obtain
\begin{equation}
	\label{hprime_bound}
	\left|\frac{H'}{H}\right| \geq \frac{1}{\sqrt{d-2}},
\end{equation}
where we have made use of Equation~\eqref{einstein2}, which implies $\dot H \leq 0$. With $\dot H = H' \dot\phi$, and assuming $\dot\phi > 0$ (the opposite case $\dot\phi < 0$ leads to the same conclusion), it follows that $H' < 0$. The inequality therefore becomes
\begin{equation}
	\label{hprime_inequality}
	-\frac{H'}{H} \geq \frac{1}{\sqrt{d-2}}.
\end{equation}
Upon integration, this yields
\begin{equation}
	\label{hubble_bound}
	H(\phi) \leq H(\phi_i) \, e^{-\frac{1}{\sqrt{d-2}}(\phi - \phi_i)}.
\end{equation}
Furthermore, the first Friedmann equation~\eqref{einstein1} implies $V \leq \frac{(d-1)(d-2)}{2} H^2$, which gives
\begin{equation}
	\label{potential_bound}
	V(\phi) < \frac{(d-1)(d-2)}{2} H(\phi_i)^2 \, e^{-\frac{2}{\sqrt{d-2}}(\phi - \phi_i)}.
\end{equation}
In the asymptotic limit of field space, this inequality translates into the following bound on the potential slope:
\begin{equation}
	\label{slope_bound}
	\left|\frac{V'}{V}\right| \geq \frac{2}{\sqrt{d-2}}.
\end{equation}

This result is a well-known constraint in the swampland program and can be derived from the original TCC bound applied to exponential potentials~\cite{bedroya2020trans}. This bound also appears in the context of string compactifications~\cite{Andriot_2023,Andriot_2020}.\footnote{There are examples in these references where the exponent of the potential saturates the TCC bound and therefore appears to violate bound~\eqref{slope_bound}. However, as noted in~\cite{Rudelius_2021}, in these examples only the volume modulus and the dilaton are considered; by including other scalars, the bound may be satisfied~\eqref{slope_bound}.} 
The above bound is identical to the bound proposed in~\cite{Rudelius_2022,Rudelius_2021}. The bound is also obtained from holographic arguments in~\cite{bedroya2024holographicorigintccdistance}.

The bound~\eqref{lower_bound} yields a constraint that is qualitatively different from~\eqref{slope_bound}. In particular, from~\eqref{lower_bound} we obtain
\begin{equation}
	\label{hprime_bound2}
	-\frac{H'}{H} \geq \frac{1}{\sqrt{d-2}} \sqrt{H},
\end{equation}
which, assuming flat potentials with $H \sim \sqrt{V}$, implies
\begin{equation}
	\label{slope_bound2}
	\left|\frac{V'}{V^{5/4}}\right| \geq \frac{2}{\sqrt{d-2}} \left(\frac{2}{(d-1)(d-2)}\right)^{1/4}.
\end{equation}
While the $V^{5/4}$ scaling is unconventional in the context of standard swampland conjectures, analogous bounds have been studied in~\cite{Rudelius_2021}, where their behavior under dimensional reduction was examined.

\begin{figure}[H]
	\centering
	\includegraphics[width=0.8\linewidth]{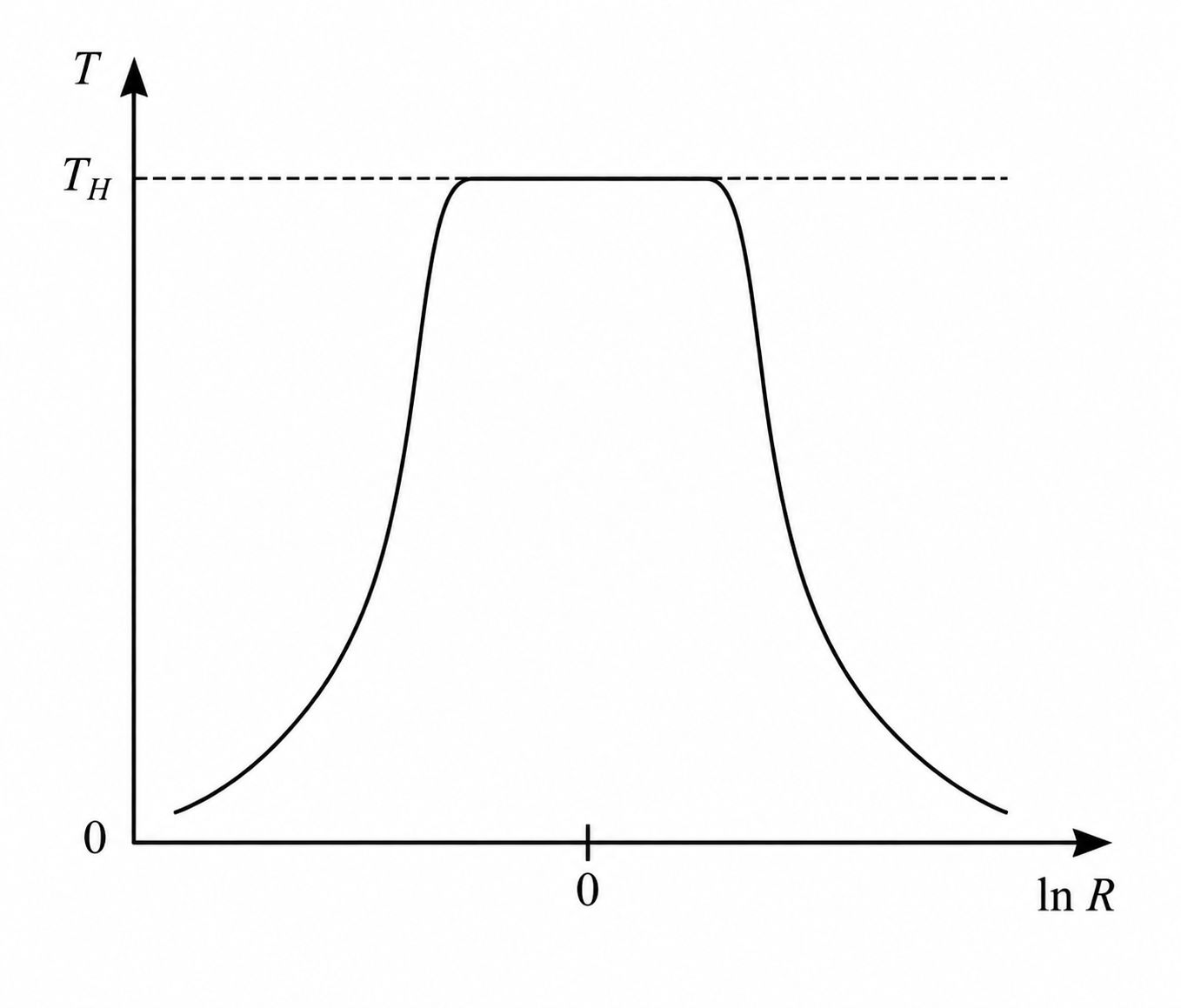}
	\caption{Temperature vs distance in string gas cosmology scenario.}
	\label{stringgas}
\end{figure}

\begin{itemize}
	\item \textbf{Second argument}: 
 \end{itemize}  
In the second argument, we use the bound on the logarithmic derivative of $\Lambda_s$ with respect to the scalar fields of the theory, $|\frac{\Lambda_s'}{\Lambda_s}|$, which we will study in the next section. The bound is given by~\cite{vandeheisteeg2023boundsfieldrangeslowly,vandeheisteeg2023boundsspeciesscaledistance}
\begin{equation}
	\label{speciesscalebound1}
	\frac{1}{\sqrt{(d-1)(d-2)}} \leq \left|\frac{\Lambda_s'}{\Lambda_s}\right| \leq \frac{1}{\sqrt{d-2}}.
\end{equation}
where the lower bound is directly obtained from conjecture~\ref{conjecture2} using the Friedmann equations, as we show in the next section. Following the emergent string conjecture, the above bound is true in the asymptotic regions of moduli space. However, the upper bound for the interior of the moduli space is also verified in some cases~\cite{vande}. The conjecture~\ref{conjecture2} implies
\bea
-\frac{\Lambda_s'}{\Lambda_s}\frac{\dot\phi}{H}>1,
\eea
using the second Friedmann equation~\eqref {fried2}, we have $\dot\phi=-(d-2)H'$, then we have
\bea
(d-2))\frac{\Lambda_s'}{\Lambda_s}\frac{H'}{H}>1,
\eea
using the upper bound in~\eqref{speciesscalebound1}, we then have
\bea
\label{Hbound}
|\frac{H'}{H}|>\frac{1}{\sqrt{d-2}},
\eea
which is the same as~\eqref{hprime_inequality} that is equivalent to $\frac{d(1/H)}{dt}>1$. In this consideration, we have used two assumptions: the first is the upper bound in the bound~\eqref{speciesscalebound1}, which is also true in the interior of the moduli space. The second assumption is the second Friedmann equation~\eqref{fried2}. The applicability of the Einstein equations in the regions where the curvature is large is under question, and the bound~\eqref{Hbound} might be corrected in these regions. However, it is not our primary goal to fully defend the bounds~\eqref{lower_bound} and~\eqref{stronger_bound}. Nonetheless, the (refined) TCC bound~\eqref{conjecture1} directly implies
\begin{equation}
	\label{horizon_bound}
	\frac{1}{H} > l_{\Lambda_s},
\end{equation}
which is naturally connected to the no-singularity conjecture, as the absence of caustics demands that $\theta = 3H$ remains finite. In certain toy models of string theory, such as string gas cosmology~\cite{Brandenberger:1988aj}, where the role of T-duality is taken into account, the temperature history of the universe is completely regular, indicating that the spacetime is regular (see Fig.~\ref{stringgas}).

 As shown in~\cite{Brandenberger:1988aj}, the temperature of the universe satisfies the relation
 \begin{equation}
 	\label{tduality_temp}
 	T(R) = T\left(\frac{1}{R}\right),
 \end{equation}
 which is a direct manifestation of T-duality. This relation implies that the temperature of the universe cannot diverge, at least in models such as string gas cosmology, in contrast to the standard big-bang scenario. The lesson to be learned is that in string theory, the temperature is bounded by the Hagedorn temperature, which is controlled by $\alpha'$, as opposed to standard FRW cosmology, and the big-bang singularity may be avoidable. Moreover, in the Hagedorn phase, the temperature is approximately constant, and we find that $a(t)$ is almost constant with $H \sim 0$. Consequently, the TCC inequality $l_{\Lambda_s} < 1/H$ is trivially satisfied.

Furthermore, given that the Kretschmann invariant is related to $\frac{d}{dt}(1/H)$, it is plausible to propose a lower bound on this derivative. We have motivated this proposal using the refined TCC bound~\eqref{conjecture2} and its implications, including Equations~\eqref{slope_bound} and~\eqref{slope_bound2}, with the latter finding support in the swampland literature~\cite{bedroya2020trans,Rudelius_2022,Rudelius_2021,Andriot_2023,Andriot_2020}.

\subsection{The refined TCC and the slope of species scale }
\label{species scale slope}
Maybe the cleanest consequence of the conjecture~\ref{conjecture2} is putting a lower bound on the logarithmic derivative of the species scale $\ Lambda_s$. In consistent effective field theories, the moduli space of scalars is generally non-compact, and there exist paths in this moduli space with infinite proper distance. On the other hand, the distance conjecture~\cite{Ooguri_2007} predicts that in the infinite distance limit, there is a tower of light states with masses given by
\begin{equation}
	\label{distance_conjecture}
	m(\phi) = m(\phi_0) \, e^{-\alpha \, d(\phi,\phi_0)},
\end{equation}
where $d(\phi,\phi_0)$ denotes the geodesic distance in the moduli space and $\alpha$ is an $\mathcal{O}(1)$ positive constant. The emergent string conjecture~\cite{lee2022} claims that in the infinite-distance limit, we encounter either a weakly coupled string theory or a decompactification phase. One may then conclude that~\cite{vandeheisteeg2023boundsfieldrangeslowly,vandeheisteeg2023boundsspeciesscaledistance}
that
\begin{equation}
	\label{alpha_bound}
	\sqrt{\frac{D-2}{(D-d)(d-2)}} \geq \alpha \geq \frac{1}{\sqrt{d-2}},
\end{equation}
where $d$ is the dimension of the spacetime and $D$ is the dimension of the higher-dimensional theory. Due to these light towers, the effective quantum gravity cutoff, known as the species scale, also depends on the scalars of the theory and runs as follows in the moduli space:
\begin{equation}
	\label{species_scale_running}
	\Lambda_s(\phi) = \Lambda_s(\phi_0) \, e^{-\lambda \, d(\phi,\phi_0)},
\end{equation}
with
\begin{equation}
	\label{lambda_bound}
	\sqrt{\frac{D-d}{(D-2)(d-2)}} \leq \lambda \leq \frac{1}{\sqrt{d-2}},
\end{equation}
where the lowest lower bound on $\lambda$ obtained when $D = d + 1$. Therefore, we obtain the following bounds on the slope of the species scale cutoff:
\begin{equation}
	\label{speciesscalebound}
	\frac{1}{\sqrt{(d-1)(d-2)}} \leq \left|\frac{\Lambda_s'}{\Lambda_s}\right| \leq \frac{1}{\sqrt{d-2}}.
\end{equation}
It is conjectured that this bound is valid everywhere in the moduli space, not just asymptotically, and evidence for its validity has been provided in several cases~\cite{vande} (at least for the upper bound). As a concrete example, the species scale in ten-dimensional Planck units for Type IIA string theory is given by~\cite{vande}
\begin{equation}
	\label{species_IIA}
	\Lambda_s = \frac{1}{(2\pi)^{1/8}} \left( \frac{3\zeta(3)}{\pi^2} e^{-3\phi/2} + e^{\phi/2} \right)^{-1/6}.
\end{equation}
In the weak-coupling limit, the species scale reduces to the string mass scale, $\Lambda_s \sim 1/l_s$, as expected. In the strong-coupling limit, Type IIA theory lifts to M-theory, and we have
\begin{align}
	l_{p,11} &= g_s^{1/3} l_s, \label{mtheory_relations} \\
	l_{p,10}^8 &= l_s^8 g_s^2, \label{planck_relations}
\end{align}
so that the species scale becomes $\Lambda_s \sim e^{-\phi/12}$, which is precisely the eleven-dimensional Planck mass. The bound~\eqref{speciesscalebound} is satisfied for the above species scale. Further examples can be found in~\cite{vande}. The lower bound in inequality~\eqref{speciesscalebound} can be easily derived from the second conjecture~\ref{conjecture2}. From the conjecture, we have
\begin{equation}
	\label{scale_factor_bound}
	\frac{a_f}{a_i} \leq \frac{\Lambda_s(t_i)}{\Lambda_s(t_f)}.
\end{equation}
Taking $t_f = t_i + \delta t$ and expanding to first order in $\delta t$, we obtain
\begin{equation}
	\label{hubble_species_relation}
	H \leq -\frac{\dot\Lambda_s}{\Lambda_s} = -\frac{\Lambda_s'}{\Lambda_s} \dot\phi.
\end{equation}
As in the previous section, we assume $\dot\phi > 0$. Therefore, we have
\begin{equation}
	\label{species_slope_bound_derivation}
	\left|\frac{\Lambda_s'}{\Lambda_s}\right| \geq \frac{H}{|\dot\phi|} \geq \frac{1}{\sqrt{(d-1)(d-2)}},
\end{equation}
where in the last inequality we have used the first Friedmann equation~\eqref{einstein1}, together with the assumption that $V$ is positive. This consequence of conjecture~\ref{conjecture2} provides strong support for its plausibility.
\section{Conclusion}
In this paper, we have studied refined versions of the TCC, namely conjectures~\ref{conjecture1} and~\ref{conjecture2}, which are formulated in terms of the species scale rather than the Planck mass. A primary consequence of these bounds is a shorter lifetime for de Sitter spacetime, which distinguishes them from the standard TCC. Moreover, conjecture~\ref{conjecture1} also leads to a bound on the cosmological constant. We have also shown that conjecture~\ref{conjecture2} violates some decelerating expansions as opposed to the standard TCC bound. From a different perspective, we have argued that the refined TCC bounds (as well as the standard TCC) imply the absence of caustic points in cosmological spacetimes, which we refer to as the no-singularity conjecture. We have also shown that conjecture~\ref{conjecture2} places a lower bound on $\frac{d}{dt}(1/H)$, which is sufficient to exclude de Sitter spacetime. Finally, we have demonstrated that conjecture~\ref{conjecture2} implies a lower bound on the logarithmic derivative of the species scale,
\begin{equation}
\left|\frac{\Lambda_s'}{\Lambda_s}\right| \geq \frac{1}{\sqrt{(d-1)(d-2)}},
\end{equation}
which is consistent with other swampland conjectures.

\begin{acknowledgments}
I wish to thank Mahdi Torabian and Reza Ghomi Shurkai for useful discussions and comments. The author also acknowledges the use of AI tools for grammatical polishing and for generating Fig.~\ref{stringgas}.
\end{acknowledgments}

	\bibliographystyle{apsrev4-1}
	\bibliography{reference}  
\end{document}